\documentclass[%
 reprint,
superscriptaddress,
 amsmath,amssymb,
 aps,
]{revtex4-1}

\usepackage{graphicx}
\usepackage{dcolumn}
\usepackage{bm}

\begin{document}

\title{Small-World Propensity in Weighted, Real-World Networks}

\author{Sarah Feldt Muldoon}\affiliation{Department of Bioengineering, University of Pennsylvania, Philadelphia, PA 19104 USA}\affiliation{US Army Research Laboratory, Aberdeen Proving Ground, MD 21005, USA}
\author{Eric W. Bridgeford}\affiliation{Department of Bioengineering, University of Pennsylvania, Philadelphia, PA 19104 USA}\affiliation{Department of Biomedical Engineering, Johns Hopkins University, Baltimore, MD 21218 USA}
\author{Danielle S. Bassett}\affiliation{Department of Bioengineering, University of Pennsylvania, Philadelphia, PA 19104 USA}\affiliation{Department of Electrical and Systems Engineering, University of Pennsylvania, Philadelphia, PA 19104 USA}\affiliation{Corresponding author: dsb@seas.upenn.edu}



\begin{abstract}
Quantitative descriptions of network structure in big data can provide fundamental insights into the function of interconnected complex systems. Small-world structure, commonly diagnosed by high local clustering yet short average path length between any two nodes, directly enables information flow in coupled systems, a key function that can differ across conditions or between groups. However, current techniques to quantify small-world structure are dependent on nuisance variables such as density and agnostic to critical variables such as the strengths of connections between nodes, thereby hampering accurate and comparable assessments of small-world structure in different networks.  Here, we address both limitations with a novel metric called the Small-World Propensity (SWP). In its binary instantiation, the SWP provides an unbiased assessment of small-world structure in networks of varying densities. We extend this concept to the case of weighted networks by developing (i) a standardized procedure for generating weighted small-world networks, (ii) a weighted extension of the SWP, and (iii) a stringent and generalizable method for mapping real-world data onto the theoretical model. In applying these techniques to real world brain networks, we uncover the surprising fact that the canonical example of a biological small-world network, the \emph{C. elegans} neuronal network, has strikingly low SWP in comparison to other examined brain networks. These metrics, models, and maps form a coherent toolbox for the assessment of architectural properties in real-world networks and their statistical comparison across conditions.
\end{abstract}

\keywords{small-world networks | weighted networks | brain connectivity | network structure | network generation}
\maketitle

\section{Introduction}
In the era of big data sciences, the ability to infer fundamental principles of system function from large sets of interconnected variables is of increasingly pressing importance for structural characterization, functional prediction, and novel network design. Traditionally, network science has been posited as a particularly appealing framework in which to form such inferences based on its simplicity and ubiquitous utility \cite{Newman2010}. Yet, it is exactly this simplicity -- arguably network science's greatest strength -- which often ignores critical system details, causing it to dual as the field's greatest potential weakness \cite{Proulx2005}.  For example, many network-based tools were originally developed to examine isolated networks constructed from binary links.  However, storage facilities around the world now house rich new data offering multiple instances of the same or related networks, with finely measured weights on links between nodes. Progress in using these data to understand and controllably manipulate complex systems requires complementary theoretical advances in the realism of network-based analysis tools.

Particularly relevant tools are those used for characterization: quantitative statistics to describe the organization or structure of networks, from which one can infer properties of their function. A quintessential example is the structure of small-worldness, ubiquitous across many real-world networks \cite{Costa:2011do}, whose local clustering combined with the ability to move quickly through the network has been shown to have important implications for functions from synchronizability to information flow \cite{Nishikawa:2003wy, Lu:2004uh, Roxin:2004th, Oliveira:2014dx}.  Watts and Strogatz formalized the concept of a small-world network by describing a simple theoretical model on binary networks \cite{Watts:1998vz}.  Nodes are placed on a lattice and connected to their nearest neighbors within a radius, $r$, giving the network a high clustering coefficient.  Shortcuts are then introduced to the network by randomly rewiring each edge with some probability, $p$.   For $p=0$ the network is a lattice, for $p=1$ the network is random, and for a range of intermediate $p$, the network is small-world.  

While this model has proven useful for theoretical investigations \cite{Percha:2005tv, Netoff:2004ub, Newman:1999cv, Newman:1999wa}, it does not provide a direct metric to assess small-worldness in observed real-world networks. If one has observations of a network over a range of size scales, one can determine small-worldness by asking whether the characteristic path length scales as the $log(N)$ where $N$ is the number of nodes in the network \cite{Newman:2000hc}.  More commonly, however, one has observations of networks at a single size scale, and therefore needs to map real-world data to the theoretical Watts-Strogatz model.  Such a mapping can be achieved by a comparison of the observed clustering coefficient and path length to that of random and/or lattice networks \cite{Humphries:2008bx,Humphries:2006jy,Telesford:2011dv}.  

Critically, these models -- and associated tools and mappings -- have lagged behind advances in the data sciences enabling more comprehensive network measurement. Specifically, these statistics are dependent on network density and neglect critical variables such as the strengths of connections between nodes, limiting their ability to diagnose and compare small-world structure in different networks. To address these limitations, we introduce a novel diagnostic called the Small-World Propensity (SWP), which quantifies the extent to which a network displays small-world characteristics while accounting for variation in network density. We next present (i) a standardized procedure for generating weighted small-world networks, (ii) a weighted extension of the SWP, and (iii) a stringent and generalizable method for mapping real-world data onto the theoretical model.  We verify our methods using standard benchmark networks and use them to examine small-world properties of real-world brain networks in humans and neuronal networks in worms.  Surprisingly, we observe that the neuronal network of \emph{C. elegans}, originally touted as a quintessential example of a biological small-world network, shows especially weak SWP.

\section{Results}

\subsection{Small-World Propensity}
To quantify the extent to which a network displays small-world structure, we define the Small-World Propensity, $\phi$, to reflect the deviation of a network's clustering coefficient, $C_{obs}$, and characteristic path length, $L_{obs}$, from both lattice ($C_{latt}$, $L_{latt}$) and random ($C_{rand}$, $L_{rand}$) networks constructed with the same number of nodes and the same degree distribution:

\begin{equation}
\label{eq:SWP}
\phi = 1-\sqrt{\frac{\Delta_{C}^{2}+\Delta_{L}^{2}}{2}},
\end{equation}
where
\begin{equation}
\Delta_{C}=\frac{C_{latt}-C_{obs}}{C_{latt}-C_{rand}}
\end{equation}
and
\begin{equation}
\Delta_{L}=\frac{L_{obs}-L_{rand}}{L_{latt}-L_{rand}}.
\end{equation}

\begin{figure}
\centerline{\includegraphics{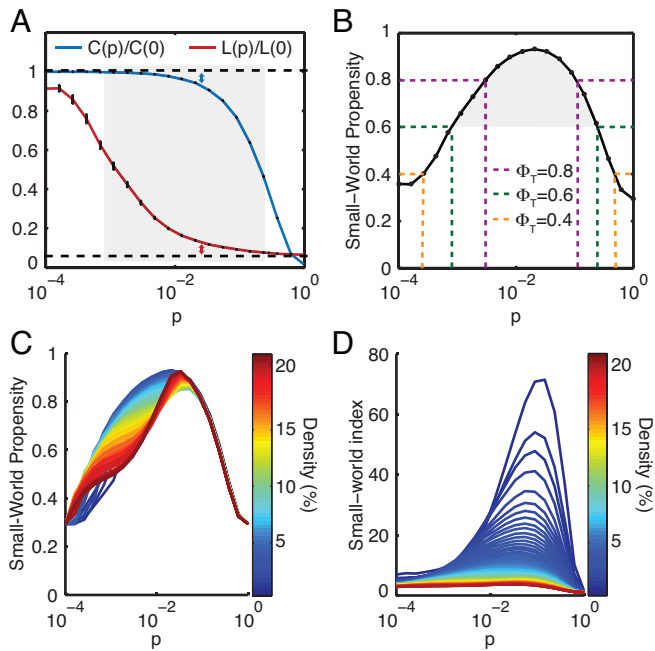}}
\caption{\textbf{Small-World Propensity in binary networks.}  \emph{(A)} Clustering coefficient and path length as a function of the rewiring parameter, $p$, for a standard Watts-Strogatz formulation of a small-world network with $N=1000$ nodes and $r=5$.  The dashed horizontal lines mark the baseline value of the clustering coefficient in a similar lattice network (top) and the baseline value for the path length in a similar random network (bottom). \emph{(B)} SWP calculated for the same network as in panel \emph{(A)}.  Error bars represent the standard error of the mean calculated over 50 simulations, and the shaded regions represent the range denoted as SW if using a threshold value of $\phi_T=0.6$. \emph{(C)} SWP as a function of network density (increasing $r$ for $N=1000$ nodes).  \emph{(D)} Small-world index for the same networks as in panel \emph{(C)}.  \label{fig1}}
\end{figure}

\noindent The ratio $\Delta_{C/L}$ represents the fractional deviation of the metric ($C_{obs}$ or $L_{obs}$) from its respective null model (a lattice or random network).\footnote{Because it is possible for real-world networks to display path lengths or clustering coefficients that exceed that of a lattice or random network, we bound these values between $0$ and $1$. Thus, if $\Delta_{C/L} > 1$ we set $\Delta_{C/L}=1$ and if $\Delta_{C/L} < 0$ we set $\Delta_{C/L}=0$, which guarantees that $\phi$ is bounded in the range $[0\: 1]$.} See arrows in Fig.~\ref{fig1}A for a schematic of the deviation, and see \emph{Materials and Methods} for mathematical definitions of $C$ and $L$. 

In a standard Watts-Strogatz model, we observe that SWP is maximal for network configurations with the greatest small-world characteristics (Fig.~\ref{fig1}A-B).  For small $p$, the SWP is low, driven by a high path length compared to that of a random network (resulting in a large $\Delta_{L}$).  However, as $p$ increases, the path length quickly becomes closer to that of a random network, while the network retains a high clustering coefficient, similar to a lattice.  Thus, when the SWP is maximal for $p\approx0.02$, we see equal contributions from $\Delta_{C}$ and $\Delta_{L}$.  As $p$ increases further, the network becomes increasingly random: the path length remains small, and the low SWP is now driven by the lack of local clustering (high $\Delta_{C}$).

As illustrated by the application to the Watts-Strogatz model, the SWP is best utilized as a comparative metric to describe the extent to which a network displays small-world structure. While a definitive hard threshold on this value is meaningful only in the context of a theoretical model \cite{Telesford:2011dv}, we pragmatically choose a reference value of $\phi_T=0.6$ to distinguish a network with a strong small-world propensity from a network with weak small-world propensity (Fig.~\ref{fig1}B).\footnote{It is also important to stress that although we have suggested $\phi_T=0.6$ as a potential useful threshold above which a network falls in the range typically identified as SW when applying a Watts-Strogatz characterization, more or less stringent definitions of SW structure could be chosen.}

The SWP has the critical ability to discern differences in small-world structure across network densities (Fig.~\ref{fig1}C).  As density increases, the range of values spanned by the clustering coefficient and path length decreases (Supplementary Fig.~1). The SWP is normalized by these ranges, minimizing the effects of network density on the calculation.  The utility of this feature is particularly evident in comparison to the commonly used small-world index, $\sigma$, proposed by Humphries et al. \cite{Humphries:2008bx,Humphries:2006jy}. As seen in Fig.~\ref{fig1}C-D, unlike the small-world index, the SWP retains a large dynamic range even as network density is increased, accurately pinpointing a small-world regime. (See the {\it SI} for a discussion of small-worldness in high density networks.)

\subsection{Generating Weighted Small-World Networks}

To extend the SWP to weighted networks, it is necessary to define models of weighted lattice and random networks.  Here, we take inspiration from the importance of physical distance in spatially embedded networks \cite{Barthelemy:2011dq}, and the observation that in many real physical and biological networks, the strength of an edge is inversely correlated with the physical distance between nodes, i.e., nodes located near one another tend to be linked by stronger edges than nodes located far from one another \cite{Kaiser2006,Levy2010,Bassett2015}.  By incorporating physical space, we define a re-wiring mechanism that allows a weighted network to be manipulated from a lattice network to a random network while maintaining the distribution of edge strengths:

\begin{enumerate}
\item{Begin with a network of $N$ nodes that are arranged on a lattice and connected to all neighbors within a radius, $r$.}
\item{Assign edge weights $w_{ij}$ according to the distance between nodes $d_{ij}$:
\begin{equation}
w_{ij}=D_{max}-d_{ij}
\end{equation}
where $D_{max}=max\{d_{ij}\}+\delta$ is the maximum possible distance between two nodes plus a single unit of measurement of the lattice spacing.\footnote{For a lattice reflecting real data, $\delta$ would equal the precision of the measurement.}$^{,}$\footnote{The inclusion of $\delta$ ensures that no edges will be assigned a weight of $0$.}}
\item{Re-wire each edge with probability, $p$, retaining the weight of the edge.}
\end{enumerate}

\begin{figure}
\centerline{\includegraphics{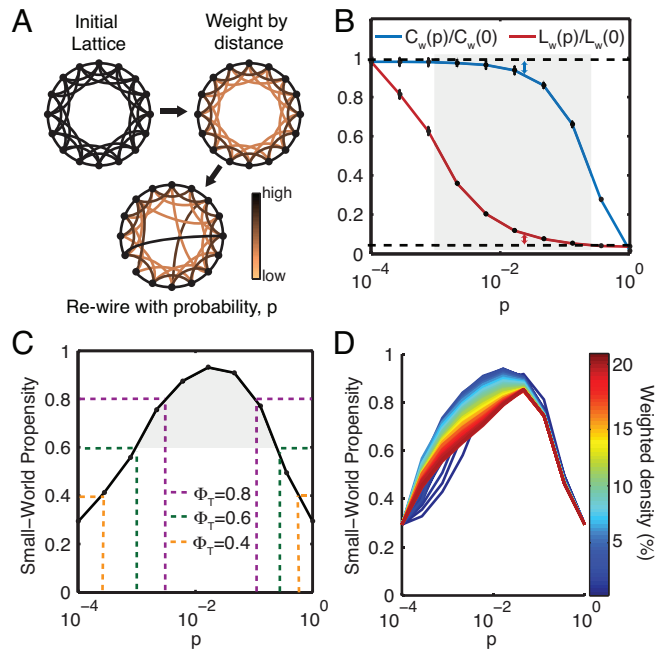}}
\caption{\textbf{Small-World Propensity in weighted networks.}  \emph{(A)} Generation of weighted small-world networks.  After building a lattice, the edges are weighted by distance such that close edges have a higher strength than distant edges.  These edge weights are then retained as links are rewired with a probability, $p$,  to create a weighted small-world network.  \emph{(B)} Weighted clustering coefficient and weighted path length as a function of the rewiring parameter, $p$, for a weighted formulation of a small-world network with $N=1000$ nodes and $r=5$. \emph{(C)} Weighted SWP calculated for the same network as in panel \emph{(B)}.  Error bars represent the standard error of the mean calculated over 50 simulations, and the shaded regions represent the range denoted as small-world if using a threshold value of $\phi_T=0.6$.  \emph{(D)} Weighted SWP as a function of network density (increasing $r$ for $N=1000$ nodes).  
\label{fig2}}
\end{figure}

Following network formation, we assess small-world structure as a function of the rewiring probability by computing weighted versions of the clustering coefficient, $C_{w}$, and characteristic path length, $L_{w}$, (see Fig.~\ref{fig2}B and \emph{Materials and Methods} for definitions).  These variables monitor a transition of our weighted networks through the small-world regime in a manner that is highly similar to that observed in the original Watts-Strogatz procedure for binary networks (compare Figs.~\ref{fig1}A \& \ref{fig2}B). We quantify this transition by computing a weighted version of the SWP using $C_{latt}=C_{w}(p=0)$ and $L_{rand}=L_{w}(p=1)$. We observe that the weighted SWP behaves similarly to the unweighted SWP (compare Fig.~\ref{fig1}B \& Fig.~\ref{fig2}C), and maintains a large dynamic range as network density is increased (Fig.~\ref{fig2}D).

\subsection{Mapping Real-World Observations to Theory}
To compute the SWP in a real-world network, it is necessary to determine the appropriate comparable lattice and random networks. To account for the effects of network density, we construct lattice and random networks that maintain the network density -- number of nodes and edge weight distribution -- observed in the real-world network. Specifically, we construct a comparable weighted lattice by arranging the observed edge weights such that the edges that correspond to the smallest Euclidean distance between nodes are assigned the highest weights (Supplementary Fig.~2). For example, for a $1$D lattice with $N$ nodes and unit spacing between nodes, we have $N$ edges with a Euclidean distance of $d=1$ between nodes.  We therefore rank the observed edge weights by decreasing strength and randomly distribute the connections with the $N$ highest weights among the edges representing $d=1$.  We proceed to distribute the next $N-1$ edges of highest weight among the edges of the lattice corresponding to $d=2$, and we continue to proceed in this manner until the total number of edges in the real-world network have been placed in the lattice. To create a comparable random network, the observed edge weights are randomly distributed among the $N$ nodes of the network.  These reference networks are then used to calculate the SWP of the observed network.

\subsection{Synthetic Benchmark Networks}
\begin{figure}
\centerline{\includegraphics{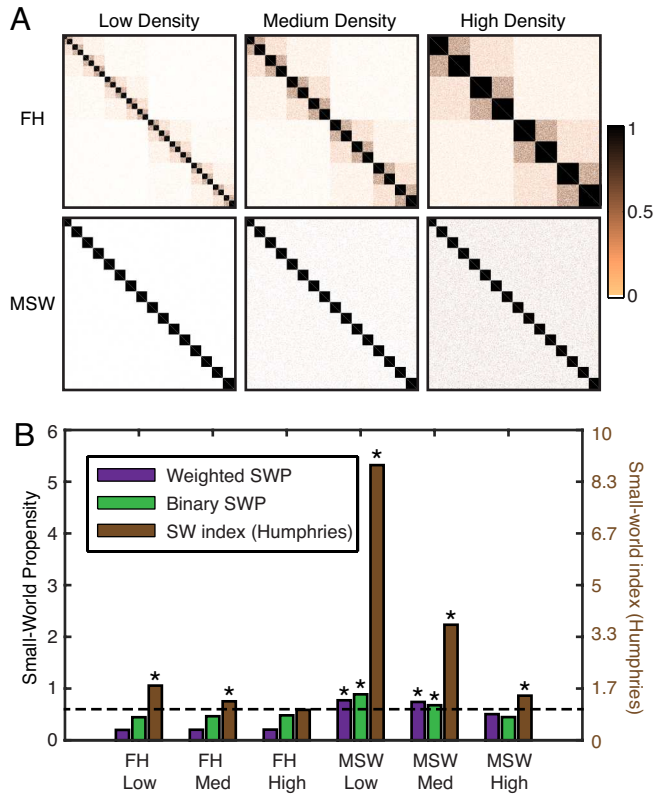}}
\caption{\textbf{Benchmark networks.}  \emph{(A)} Adjacency matrices depicting network organization for increasingly dense Fractal Hierarchical (FH) and Modular Small-World (MSW) Networks. \emph{(B)} Comparison of the unweighted SWP, weighted SWP, and small-world index when applied to the matrices in panel \emph{(A)}.  The dashed line represents the chosen threshold for indicating small-world structure ($\phi_T=0.6$ for SWP or $\sigma_T=1$ for the small-world index).  Stars denote networks classified as small-world.  \label{fig3}}
\end{figure}

To further assess the validity of the SWP, we examine its performance on weighted networks with well studied structure: fractal hierarchical (FH) and modular small-world (MSW) networks (see \emph{Materials and Methods}). While both types of networks have a modular structure and therefore dense local clustering, the networks differ in terms of their path length: the MSW networks incorporate random shortcuts of moderate connection strength between modules, decreasing the weighted path length, while the FH networks are composed of hierarchically organized modules that are weakly interconnected, increasing the weighted path length (see Fig.~\ref{fig3}A).  It follows that MSW networks have greater small-world structure than FH networks.  

In Fig.~\ref{fig3}, we compare the ability of three metrics to detect small-world structure in these networks constructed at various densities: (i) the weighted SWP computed on the true networks, and (ii) the un-weighted SWP and (iii) Humphries' small-world index computed on binarized versions of the true networks.  In the FH networks, we observe that both the weighted and unweighted SWP values are relatively independent of density, while the small-world index is strongly dependent on density. Moreover, the small-world index classifies low and medium density FH networks as having SW properties ($\sigma > 1$), despite the long path lengths characteristic of these networks.  The SWP instead reflects our expectations, and remains below the small-world threshold ($\phi_T=0.6$) even in the case of binarized networks. Importantly, when edge weights are retained, the SWP reveals that the FH networks are quite different than their MSW counterparts and do not show a small-world structure. 

In the MSW networks, we observe that the SWP values are sensitive to a meaningful change in the network structure as density increases.  Specifically, the path length becomes closer to that of a random network, but the addition of shortcuts between modules reduces the clustering coefficient thereby increasing the deviation from the clustering coefficient of a comparable lattice network $\Delta_C$. This change is reflected by decreasing SWP.  The small-world index is less sensitive to the deviation in clustering, and classifies all MSW networks as having small-world structure.

\subsection{Real-World Brain and Neuronal Networks}
\begin{figure}
\centerline{\includegraphics{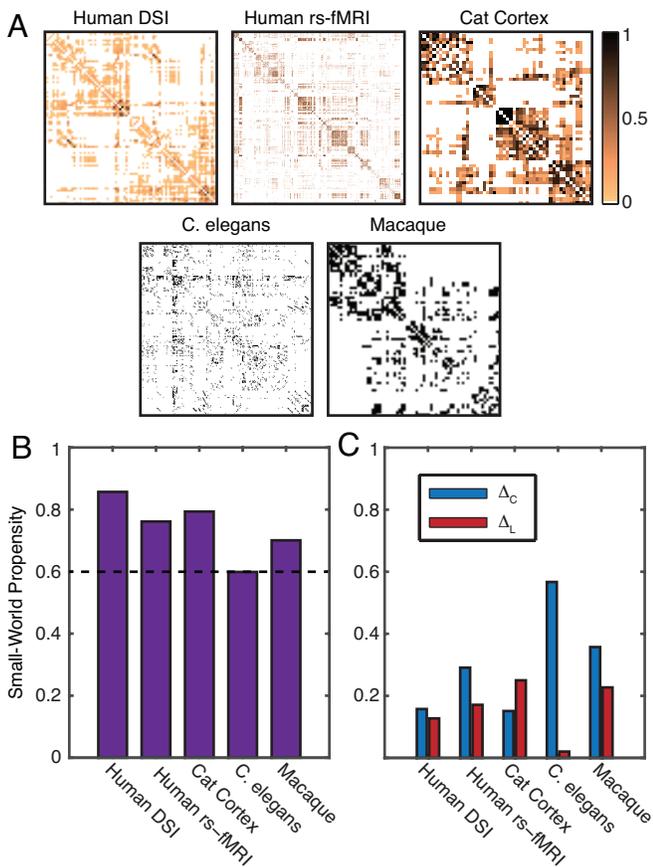}}
\caption{\textbf{Real-world brain networks.}  \emph{(A)} Adjacency matrices for three weighted brain networks (upper) and two binary brain networks (lower).  All matrices are symmetric.  \emph{(B)} SWP for the matrices shown in panel \emph{(A)}. The dashed line denotes the $\phi_T=0.6$ threshold. \emph{(C)}  Breakdown of the SWP into the individual contributions from the clustering coefficient and path length.  High values of $\Delta_{C}$ or $\Delta_{L}$ indicate a large deviation from the comparable benchmark value, which results in a reduction of the SWP.  
\label{fig4}}
\end{figure}

The quintessential example of a small-world network in biology is the neuronal network of \emph{C. elegans} \cite{Watts:1998vz}. Indeed, brain networks more broadly have been described as small-world for the past decade \cite{Bassett:2006kc,Stam:2012kk}. Yet, with advanced imaging and methodological techniques that provide more detailed measurements of these data with varying network densities and fine-scale estimates of edge strength, the assumptions of small-worldness have been brought into question \cite{Hilgetag2015}. Here we aim to resolve this controversy by applying the SWP to a representative set of  structural and functional brain networks from several species whose properties have been well-studied in previous literature, and whose edges are both binary and weighted (Fig.~\ref{fig4}A).  The three weighted networks are (i) a structural network representing the number of white matter tracts connecting 83 brain regions obtained from human diffusion spectrum imaging (DSI) data \cite{Gu:2014vx}, (ii) a functional network given by correlations between the blood oxygen level dependent signal of 638 brain regions measured using resting state functional magnetic resonance imaging (rs-fMRI) \cite{Crossley:2013kl}, and (iii) a structural network of the cat cortex obtained from tract-tracing studies between 52 brain regions \cite{Scannell:1999tg}. The two binary networks are (i) the neuronal network of \emph{C. elegans} representing synaptic and gap junction connections between 279 neurons \cite{Varshney:2011ju}, and (ii) a structural network derived from tract-tracing studies between 71 regions in the macaque cortex \cite{Young:1993if}. 

We observed that all networks displayed relatively large SWP values, indicating that brain and neuronal networks in general do indeed display small-world properties (Fig.~\ref{fig4}B).  Surprisingly, the network with the lowest SWP, sitting just below the $\phi_T=0.6$ threshold, is the neuronal network of \emph{C. elegans}, the quintessential example of a small-world network in biology \cite{Watts:1998vz}. We explored this surprising result by examining the contributions of the clustering ($\Delta_C$) and the path length ($\Delta_L$) to the SWP value (Fig.~\ref{fig4}C).  In contrast to the other brain networks studied, the neuronal network of \emph{C. elegans} displays drastically different contributions: a remarkably high $\Delta_C$ indicating divergence from a lattice network, and an exceptionally low $\Delta_L$ indicating similarity to a random network. The observed SWP is therefore driven almost entirely by the short path length, indicating that, in reality, this canonical example of a small-world network, does not strongly embody small-world principles.

\section{Discussion}
As petabytes of rich new data quickly file onto growing storage farms, complex interconnected systems are being increasingly queried.  Using these data to characterize and manipulate systems requires the development of network-based analysis tools that accurately account for the newly measured features of the data.  Here we offer a novel network statistic (SWP) that accurately quantifies small-world structure in networked systems, that is agnostic to nuisance variables such as density and exquisitely sensitive to critical variables such as the strengths of connections between nodes. This enterprise has required the development of (i) a standardized procedure for generating weighted small-world networks, (ii) a weighted extension of the SWP, and (iii) a stringent and generalizable method for mapping real-world data onto the theoretical model. We have illustrated the application and utility of these methods in the context of both benchmark and real-world data. The work represents a single effort in the much larger space of meeting the demands of big data with increasingly sophisticated network-based tools.  

\subsection{Small-World Statistics for Real-World Data}

The small-world index \cite{Humphries:2008bx,Humphries:2006jy} is the most common statistic to quantify small-world structure in binary networks, but produces values greater than one for a large range of network topologies, obscuring accurate interpretations.  Other measures monitor binary topologies with greater sensitivity but remain dependent on network density and independent of edge weights \cite{Telesford:2011dv}. More recent work has used weighted clustering coefficients and path lengths to compare observed networks to random networks \cite{Li:2007gf}, and to measure local small-world structure \cite{Bolanos:2013jd} and topological dimension. However, the generalizability of these methods has been hampered by the lack of a theoretical model to construct weighted small-world networks and to study the transition in and out of the small-world regime.  (Although, see \cite{Zhang:2010iq} for a method to generate weighted scale-free small-world networks.)  Our work directly fills this gap, allowing the creation of weighted small-world networks with varied degrees of SWP that can be used to compare and contrast network behavior and dynamics throughout the small-world regime.

\subsection{Binary Categories vs. Continuous Narration}

Recent network-focused efforts in the applied mathematics, physics, computer science, and engineering communities evidence the age-old tension to retain model simplicity while maximizing pragmatic utility. Network statistics are no exception: we often wish to obtain binary categorizations rather than continuous descriptions. For example, one might wish to emphatically state that a network does or does not display community structure, rich-club architecture, or small-worldness. Yet, arguably a more interesting and useful statement might assess gradations of these properties in real-world systems \cite{Rombach2014,Bassett2013}.  The true power of the SWP lies not in its ability to define small-world structure, but instead in its ability to quantify and compare the \emph{continuous degree} of small-world structure between different networks. Our work therefore complements ongoing efforts to extend traditionally categorical distinctions to continuous measurments in the context of core-periphery structure \cite{Rombach2014} and community structure \cite{Palla2005}. 

\subsection{Pragmatic Utility in Neuronal and Brain Networks}

Network statistics that are independent of density are critical for network comparisons in real-world settings. In the context of brain networks, the need for such statistics is underscored by the growing interest in examining neuronal networks across (i) time in development and normal aging \cite{Shi:2013eb,Tsujimoto:2008ha}, (ii) health and disease \cite{Hoffman:1997tm,vandenHeuvel:2010ka}, and (iii) different stages of neurodegeneration \cite{Lo:2010kz}. In these and similar contexts, it becomes very difficult to determine if observed differences in small-world structure are simply a function of the differences in network density, or if they represent a true form of topological reorganization. By minimizing the effects of network density in the computation of the SWP, we allow for a more direct comparison of the topological network structure across time, and between groups.

Network statistics that are highly sensitive to edge weights provide increased sensitivity to network function. In the context of brain networks, the need for such statistics has only recently been actively appreciated, as a growing body of literature demonstrates that both healthy and diseased brain function is differentially driven by strong \emph{versus} weak connections \cite{Bassett:2012dh,Khambhati:2014wv}.   Weak connections have traditionally been ignored because of commonly applied thresholding techniques \cite{Bassett:2012dh}, but have recently been identified as potential biomarkers in psychiatric pathologies \cite{Bassett:2012dh} and as predictors of cognitive function and fluid intelligence \cite{Schneidman:2006he,Santarnecchi:2014ju,Cole:2012cd}. In the future, it will be interesting to build on weighted network diagnostics like the SWP to provide novel quantifications of weak connectivity and its role in cognition. 

\subsection{Surprising Biological Insights}
The canonical example of a biological small-world network is the wiring diagram of \emph{C. elegans}. However, the observation of small-world structure in this organism has been built on a simplification of the weighted wiring diagram to a binary graph. Using the weighted SWP, we observe that in fact, this network displays very little small-world propensity, predominantly due to a lack of local clustering. We speculate that several key biological feautres of this network may explain this surprising contradiction to the historical literature.  The wiring diagram of \emph{C. elegans} is (i) a micro-scale network, (ii) represents neuron-to-neuron connections throughout the entire body of the organism rather than only the head, and (iii) is drawn from a comparatively symplistic organanism, evolutionarily speaking \cite{Bassett:2010hf}. These features of scale, physical extent, and evolutionary class may drive toplogical properties away from the small-world architecture to enable a different class of neural functions than those associated with the brains of higher-order animals. Future work must more stringently examine other canonical examples of small-world networks in diverse real-world systems, and thereby build a more accurate assessment of the role of topology in complex system function.

\section{Materials and Methods}
All analysis was done using MATLAB (MathWorks) and code to compute the SWP in real-world networks can be accessed at  http://www.seas.upenn.edu/$\sim$dsb/.

\subsection{Clustering coefficient}
To calculate the clustering coefficient, as in \cite{Watts:1998vz}, we first calculate the local clustering coefficient for each node, $c_i$ and then define the clustering coefficient, $C$, to be the average of the local coefficients 
\begin{equation}
C=\frac{1}{N}\sum_i c_i.
\end{equation}
For binary networks, the local coefficient, $c_i$, of each node, $n_i$, is the fraction of closest neighbors of $n_i$ that are also connected.  Multiple extensions of $c_i$ to weighted networks have been proposed:

Onnela et al. \cite{Onnela:2005de}:
\begin{equation}
c_{i,O} = \frac{1}{k_i(k_i-1)}\sum\limits_{j,k}(\hat{w}_{ij}\hat{w}_{jk}\hat{w}_{ik})^{1/3},
\label{c_onnela}
\end{equation}

Barrat et al. \cite{Barrat:2004bk}:
\begin{equation}
c_{i,B} = \frac{1}{k_i(k_i-1)}\sum\limits_{j,k}\frac{w_{ij}+w_{ik}}{2\langle w_i \rangle}a_{ij}a_{jk}a_{ik},
\end{equation}

Zhang et al. \cite{Zhang:2005er}:
\begin{equation}
c_{i,Z} = \frac{\sum\limits_{j,k}\hat{w}_{ij}\hat{w}_{jk}\hat{w}_{ik}}{(\sum\limits_k\hat{w}_{ik})^2-\sum\limits_k(\hat{w}_{ik}^2)}.
\end{equation}
In these equations, $w_{ij}$ is the strength of a connection between nodes $i$ and $j$, $\hat{w}_{ij} = w_{ij}/max(w)$,  $k_i$ is the number of edges connected to $n_i$, and $a_{ij}=1$ if a connection exists between nodes $i$ and $j$ ($a_{ij}=0$ otherwise).  For all results reported in the main manuscript, we used the definition given by Onnela et al. (Eqn. \ref{c_onnela}), but similar results were obtained when using other definitions (see Supplementary Fig.~3).

\subsection{Path length}
The characteristic path length for a network is given by 
\begin{equation}
L=\frac{1}{N(N-1)}\sum_{i\ne j}d_{ij},
\end{equation}
where for a binary network, $d_{ij}$ is the shortest path between nodes $i$ and $j$.  For weighted networks, as in \cite{Newman:2001kc}, we define the distance between two nodes to be $d_{ij}=1/w_{ij}$.

\subsection{Synthetic benchmark networks} 
The synthetic benchmark networks were created using the Brain Connectivity Toolbox \cite{Rubinov:2010jd} with modifications to make networks symmetric and with weighting schemes as defined in \cite{Lohse:2014ws}.  All synthetic networks have $N=1024$ nodes. 

Fractal hierarchical networks were created from the makefractalCIJ function provided in the toolbox, with $mx\_lv=10$, $E=2$, and $sz\_cl =5, 6,$ or $7$ for low, medium, and high density networks, respectively.  This results in networks with 6, 5, or 4 hierarchical levels with a base module size of $n=32, 64$ or $128$.  At the lowest level of the hierarchy, modules are fully connected, and connections are placed within each hierarchical level with a probability $p= 2^{-l}$, where $l$ is the hierarchical level.   The weight of each connection $w_{ij}  = p_{ij}$, such that the weight of a connection is equivalent to the probability that a connection exists.  Because this method creates a directed network, the matrices were symmetrized by selecting the upper triangle of the resultant matrix and using these connections to create an undirected, symmetric network.  This procedure resulted in final networks with weighted (binary) densities of $4.5 (10.8) \%$, $9.1 (18.7) \%$, and $17.9 (31.3) \%$ for low, medium, and high density FH networks.

Modular small-world networks were created from the makeevenCIJ function provided in the toolbox, with $N=1024$, $K=65000, 100000,$ or $150000$, and $sz\_cl=6$ for low, medium, and high density networks, respectively.  This creates networks with 16 fully connected modules of size $n=64$ and $E=K-64512$ randomly distributed edges between modules.  Connection strengths within modules were set to $w_{ij}=1$ while strengths of inter-module edges were set to $w_{ij}=0.5$.  As with the FH networks, this method creates a directed network, so the matrices were symmetrized by selecting the upper triangle of the resultant matrix and using these connections to create an undirected, symmetric network.  This procedure resulted in final networks with weighted (binary) densities of $6.1 (6.2) \%$, $7.9 (9.5) \%$, and $10.3 (14.3) \%$ for low, medium, and high density MSW networks.

%

\begin{acknowledgments}
This work was supported by the Alfred P. Sloan Foundation, the John D. and Catherine T. MacArthur Foundation, the National Science Foundation through BCS-1441502 from the ENG, CISE, and SBE directorates, and the Army Research Laboratory through contract no. W911NF-10-2-0022 from the U.S. Army Research Office. The content is solely the responsibility of the authors and does not necessarily represent the official views of any of the funding agencies. 
\end{acknowledgments}

\bibliography{wswn_bib}{}

\end{document}



\title{Supplementary Material for ``Small-World Propensity in Weighted, Real-World Networks''}

\author{Sarah Feldt Muldoon}\affiliation{Department of Bioengineering, University of Pennsylvania, Philadelphia, PA 19104 USA}\affiliation{US Army Research Laboratory, Aberdeen Proving Ground, MD 21005, USA}
\author{Eric W. Bridgeford}\affiliation{Department of Bioengineering, University of Pennsylvania, Philadelphia, PA 19104 USA}\affiliation{Department of Biomedical Engineering, Johns Hopkins University, Baltimore, MD 21218 USA}
\author{Danielle S. Bassett}\affiliation{Department of Bioengineering, University of Pennsylvania, Philadelphia, PA 19104 USA}\affiliation{Department of Electrical and Systems Engineering, University of Pennsylvania, Philadelphia, PA 19104 USA}\affiliation{Corresponding author: dsb@seas.upenn.edu}


\maketitle
\section{Effects of Density on Clustering and Path Length}

As discussed in the main manuscript and shown in Supplementary Figure 1 for a Watts-Strogatz network, the range of values spanned by the path length and clustering coefficient decreases with increasing network density.  Here we define the range of clustering or path length to be the difference between its value in lattice and random networks, $1-C(p=1)/C(p=0)$ or $1-L(p=1)/L(p=0)$.  In the main manuscript, we present results showing calculations of the SWP and small-world index for network densities in the range $0-20\%$.  At a network density of $20\%$, we can see that the path length only spans approximately $30\%$ of it's original range.  Furthermore, by the point at which the density increases to $40\%$, the path length is the same in lattice and random networks.  This implies that the concept of a Watts-Strogatz small-world network is ill-defined for these networks of higher densities, and caution should be used when trying to impose this formalism on high density networks.  The effect of network density on path length and clustering further increases the need to take edge weights into account when quantifying network structure.  If a network with many weak connections is binarized, the density of the network can become quite high which influences measurements of network properties.

\section{Mapping real-world data to the theoretical model}
In order to quantify small-world structure in real-world networks, it is necessary to define a method for mapping the observed data to the theoretical model used to generate small world networks. Specifically, the calculation of the SWP relies on the generation of a comparable lattice network ($p=0$ in the theoretical models) and a comparable random network ($p=1$ in the theoretical models).  To control for network density when generating these comparable null models, we preserve the number of nodes and the distribution of of edge weights.  As described in the main text, to construct a comparable weighted lattice, we build a 1D network such that the edges that correspond to the smallest Euclidean distance between nodes are assigned the highest weights, whereas to construct a comparable random network, the observed edge weights are randomly distributed among the nodes.   In Supplementary Fig.~\ref{si_fig2}, we give an example of the resulting comparable lattice and random networks for the Human DSI data set analyzed in the main manuscript.  In the comparable lattice network, the connections with the highest strength have been distributed along the diagonal of the adjacency matrix, while in the comparable random network, connections are randomly assigned throughout the matrix.  In both cases, we ensure that the resulting adjacency matrix is free of self-connections and remains symmetric (reflecting the undirected nature of our initial network).

\section{SWP using alternative methods of weighted clustering}
Multiple methods of computing a weighted clustering coefficient have been proposed \cite{Onnela:2005de,Barrat:2004bk,Zhang:2005er}, and in the main manuscript we present results using the weighted clustering coefficient as defined by Onnela et al. \cite{Onnela:2005de} (see \emph{Matericals and Methods}).  This particular algorithm for computing the weighted clustering coefficient reflects subgraph intensity and has the advantage of being computationally efficient.  However, other definitions of a weighted clustering coefficient exist and reflect other features of weighted clustering that might be desirable in certain data sets \cite{Saramaki:2007de}.  Here, we show the transition in and out of the small-world regime for the same weighted small-world network depicted in Fig.~2 of the main manuscript, along with the corresponding SWP, calculated for the weighted clustering coefficient defined by Barrat et al. \cite{Barrat:2004bk} (Supplementary Fig.~\ref{si_fig3}A-B) and Zhang et al. \cite{Zhang:2005er} (Supplementary Fig.~\ref{si_fig3}C-D).  The resultant weighted small-world network displays a similar transition through the small-world regime when measured using these alternative weighted clustering measures, and the associated SWP is remarkably similar to that of the SWP obtained using the Onnela measure (Fig.~2B-C).


\begin{figure*}
\centerline{\includegraphics{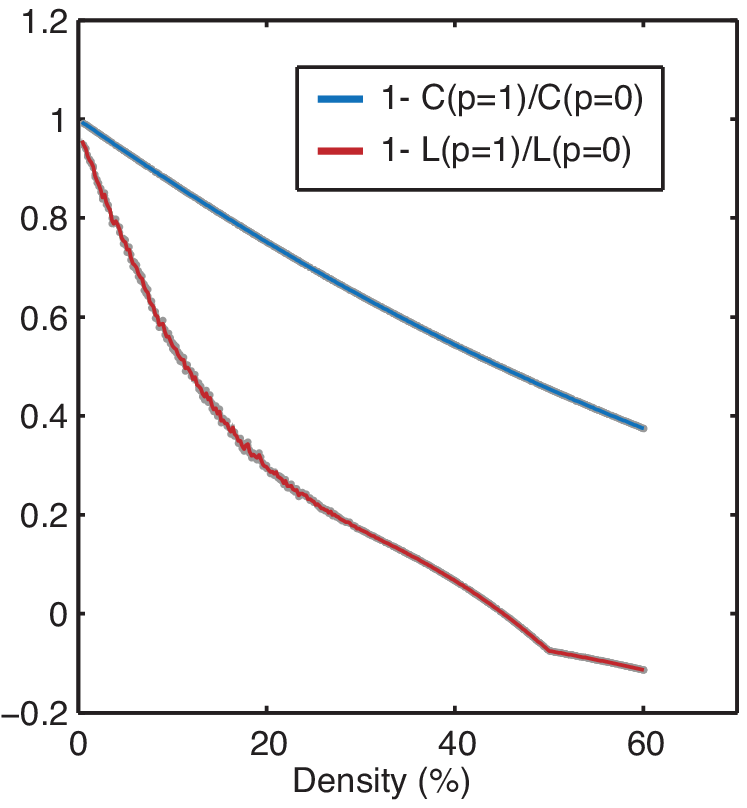}}
\caption{\textbf{Effects of density on clustering and path length} The range of the clustering coefficient (blue) and path length (red) corresponding to a Watts-Strogatz network for increasing densities (increasing $r$).  Error bars are shown in gray and represent the standard error of the mean calculated over 50 simulations.  \label{si_fig1}}
\end{figure*}

\begin{figure*}
\centerline{\includegraphics{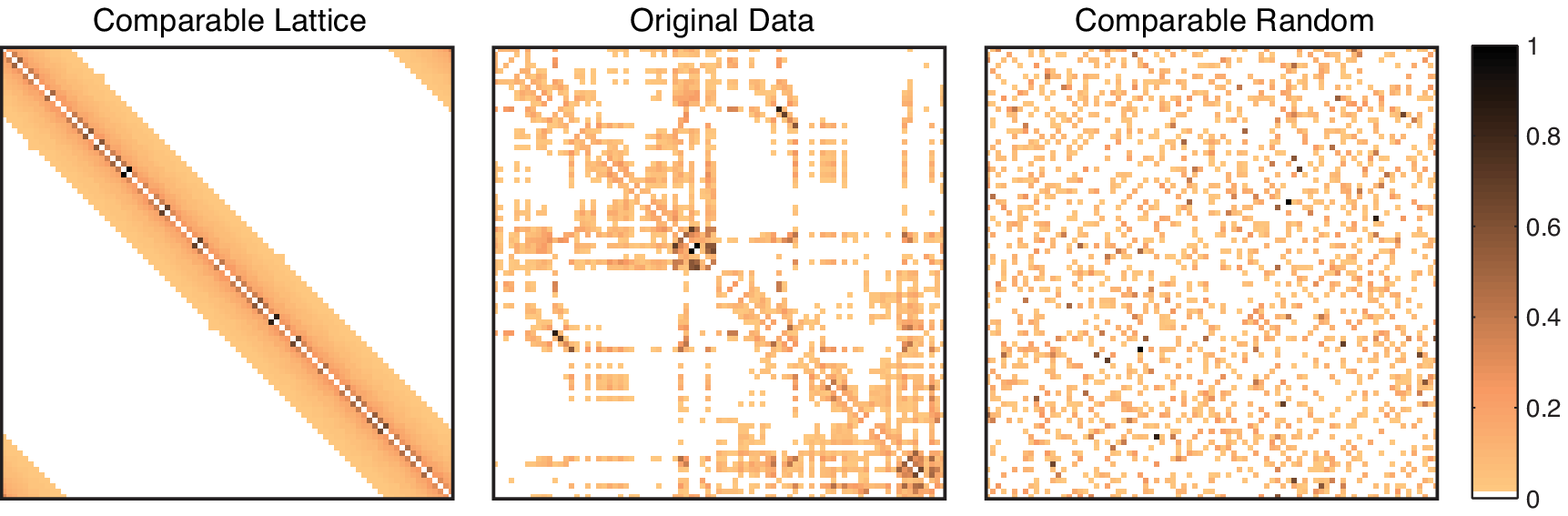}}
\caption{\textbf{Mapping real-world data to the theoretical model}  Adjacency matrices representing the mapping of the Human DSI data from the main manuscript (middle) to a comparable lattice (left) and random (right) network.  In both the comparable networks, the original edge distribution is maintained, but connections are redistributed to create a weighted lattice or random network.  \label{si_fig2}}
\end{figure*}

\begin{figure*}
\centerline{\includegraphics{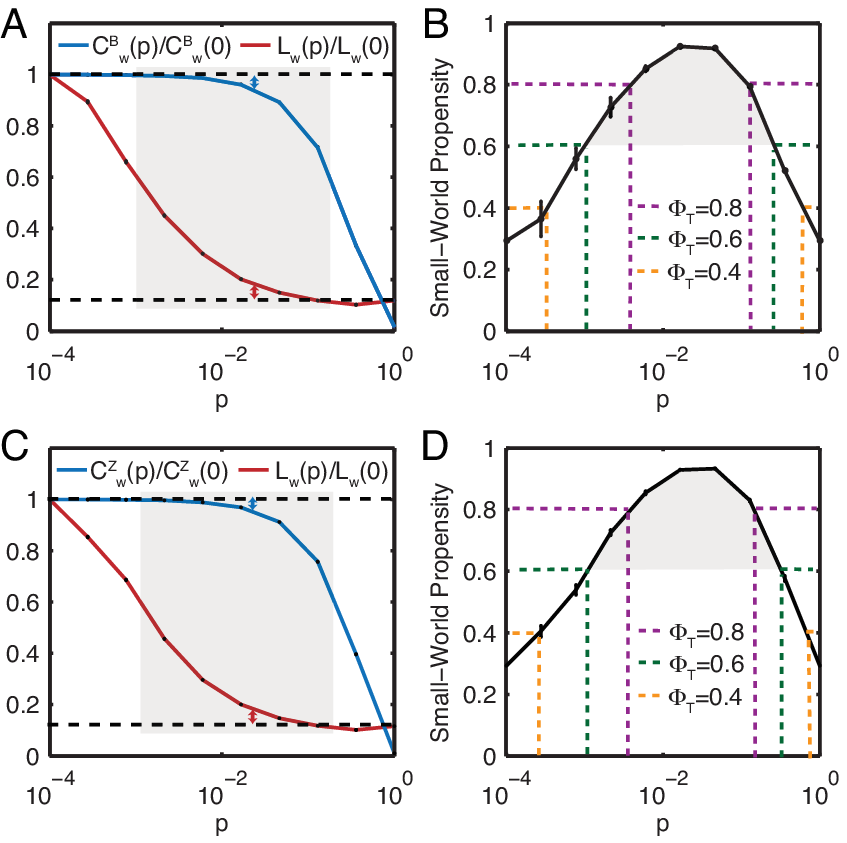}}
\caption{\textbf{SWP using alternative methods of weighted clustering}  (A) Weighted clustering coefficient as defined by Barrat et al. and weighted path length as a function of the rewiring parameter, $p$, for a weighted formulation of a SWN with $N=1000$ nodes and $r=5$.  (B) Weighted SWP calculated using the clustering in (A) . (C) Weighted clustering coefficient as defined by Zhang et al. and weighted path length as a function of the rewiring parameter, $p$, for a weighted formulation of a SWN with $N=1000$ nodes and $r=5$.  (D) Weighted SWP calculated using the clustering in (C) . Error bars represent the standard error of the mean calculated over 50 simulations, and the shaded regions represent the range denoted as SW if using a threshold value of $\phi_T=0.6$.   (A) and (C) are comparable to Fig. 2B in the main manuscript and (C) and (D) are comparable to Fig. 2C. \label{si_fig3}}
\end{figure*}



\bibliography{wswn_bib_si}{}

%
%
%
%
%
%
